\documentclass[floatfix,letterpaper,english,reprint,secnumarabic,amssymb,nobibnotes,aps,prl,superscriptaddress,nofootinbib]{revtex4-2}
\usepackage[T1]{fontenc}
\usepackage[latin9]{inputenc}
\usepackage{babel}
\usepackage{amsmath}
\usepackage{amssymb}
\usepackage{cancel}
\usepackage{graphicx}
\usepackage{wasysym}
\usepackage{color}
\usepackage[unicode=true,pdfusetitle,
 bookmarks=true,bookmarksnumbered=false,bookmarksopen=false,
 breaklinks=false,pdfborder={0 0 1},backref=false,colorlinks=false]
 {hyperref}

\usepackage{diagbox}

\makeatletter

\pdfpageheight\paperheight
\pdfpagewidth\paperwidth

\makeatother

\begin{document}
\title{$\Sigma$ Resonances from a Neural Network-based Partial Wave Analysis on $K^-p$ Scattering}

\author{Jun Shi}
\affiliation{Guangdong Provincial Key Laboratory of Nuclear Science, Institute of Quantum Matter, South China Normal University, Guangzhou51006, China }
\affiliation{Guangdong-Hong Kong Joint Laboratory of Quantum Matter, Southern Nuclear Science Computing Center, South China Normal University, Guangzhou51006, China}

\author{Long-Cheng Gui}
\affiliation{Department of Physics, Hunan Normal University, and Key Laboratory of Low-Dimensional Quantum Structures and Quantum Control of Ministry of Education, Changsha 410081, China}

\author{Jian Liang}
\email{jianliang@scnu.edu}
\affiliation{Guangdong Provincial Key Laboratory of Nuclear Science, Institute of Quantum Matter, South China Normal University, Guangzhou51006, China }
\affiliation{Guangdong-Hong Kong Joint Laboratory of Quantum Matter, Southern Nuclear Science Computing Center, South China Normal University, Guangzhou51006, China}

\author{Guoming Liu}
\email{guoming.liu@scnu.edu}
\affiliation{Guangdong Provincial Key Laboratory of Nuclear Science, Institute of Quantum Matter, South China Normal University, Guangzhou51006, China }
\affiliation{Guangdong-Hong Kong Joint Laboratory of Quantum Matter, Southern Nuclear Science Computing Center, South China Normal University, Guangzhou51006, China}

\begin{abstract}
We implement a convolutional neural network to study the $\Sigma$ hyperons 
using experimental data of the $K^-p\to\pi^0\Lambda$ reaction.
The averaged accuracy of the NN models in resolving resonances on the test data sets is 98.5\%, 94.8\% and 82.5\% for one-, two- and three-additional-resonance case.
We find that the three most significant resonances are $1/2^+$, $3/2^+$ and $3/2^-$ states with mass being 1.62(11)~GeV, 
1.72(6)~GeV and 1.61(9)~GeV, and probability being 100(3)\%, 72(24)\% and 98(52)\%, respectively,
where the errors mostly
come from the uncertainties of the experimental data.
Our results support the three-star $\Sigma(1660)1/2^+$, the one-star $\Sigma(1780)3/2^+$ and the one-star $\Sigma(1580)3/2^-$ in PDG.
The ability of giving quantitative probabilities in resonance resolving and
numerical stability make NN potentially a life-changing tool in baryon partial wave analysis,
and this approach can be easily extended to accommodate other theoretical models and/or to include more experimental data.

\end{abstract}

\maketitle

\textit{Introduction:} The study of hadron spectra helps to understand their inner structure and the underlying dynamics. 
Although hadron spectroscopy has been widely investigated by means of phenomenological models, effective theories, lattice QCD, etc., 
there still remain many uncertainties left to be clarified especially in the baryon sector: except for the ground states and the first several low-lying resonances, 
the properties and even existence of many baryon resonances are unclear~\cite{ParticleDataGroup:2022pth}.
The focus of this work is on $\Sigma$ hyperons. There are many 1-star and 2-star $\Sigma$ resonances listed in PDG, 
and further clarifying the properties of the lowest $\Sigma$ resonances has special phenomenological impacts which helps to verify different models~\cite{Zou:2007mk}. 

Partial wave analysis (PWA) of scattering reactions has been the basic approach in studies of baryon resonances. 
The general procedure of PWA is to fit experimental data with theoretical formulae containing different combinations of baryon resonances, 
and compare the $\chi^2$ of the fits in different attempts to determine their significance and properties. 
Many studies have been focused on $\Sigma$ resonances using the conventional $\chi^2$ fitting-based 
PWA~\cite{Gao:2010ve, Gao:2012zh, Zhang:2013cua, Zhang:2013sva, Kamano:2014zba, Kamano:2015hxa, Matveev:2019igl, Sarantsev:2019xxm}, 
which gains us much knowledge of $\Sigma$ spectra. However, $\chi^2$ fitting-based analysis is unable to give quantitative estimations of the statistical significance of 
possible resonances and is not stable in determining resonance's properties when more than one additional states are included. 
Regarding this, we propose the application of neural networks (NN) in PWA since it can give quantitative probabilities in category 
classification and potentially more stable in parameter regression.

NN has shown its life-changing potential in many fields of high energy and particle physics, such as handling experimental 
data produced by colliders~\cite{Radovic:2018dip,Guest:2018yhq}, extracting parton
distribution functions~\cite{Forte:2002fg,NNPDF:2021njg}, accelerating lattice QCD simulations~\cite{Shanahan:2018vcv,Kanwar:2020xzo,Favoni:2020reg},
and inferring the nature of exotic hadrons ~\cite{Liu:2022uex,Ng:2021ibr}. 
And a recent review can be
found in the summary report of Snowmass white papers~\cite{Shanahan:2022ifi}.
In general, NN is a powerful numerical tool to find out hidden relationships and correlations. 
In this sense, PWA is an ideal arena of utilizing NN since there is no direct equations relating experimental measurements and 
the resonance properties, which is kind of an analog of the famous NN application in image recognition.

$\overline{K}N\to \pi \Lambda$ scattering is a perfect channel in investigating $\Sigma$ resonances, for $\pi \Lambda$ is a pure isospin 1 channel. Currently, 
the most high-statistic and precise data of this channel is the $K^-p\to \pi^0\Lambda$ data presented by Crystal Ball collaboration for both differential 
cross sections and $\Lambda$ polarization with the center-of-mass energy from 1569 to 1676 MeV~\cite{Prakhov:2008dc}. We choose to analyze this reaction using Crystal Ball data
in the first NN application of PWA accounting for the clarity of this channel and the quality of data,
which is pertinent to show the feasibility of our new approach. 

In this article, we construct a convolutional NN to determine the quantum number with probabilities and properties of the three 
most significant $\Sigma$ resonances in the $K^-p$ scattering with sophisticated error estimation.
We demonstrate the feasibility and advantage of using NN in PWA and propose more widely application of NN in future PWA of baryon resonances.

\begin{figure}[htbp]
 \includegraphics*[width=0.4\textwidth]{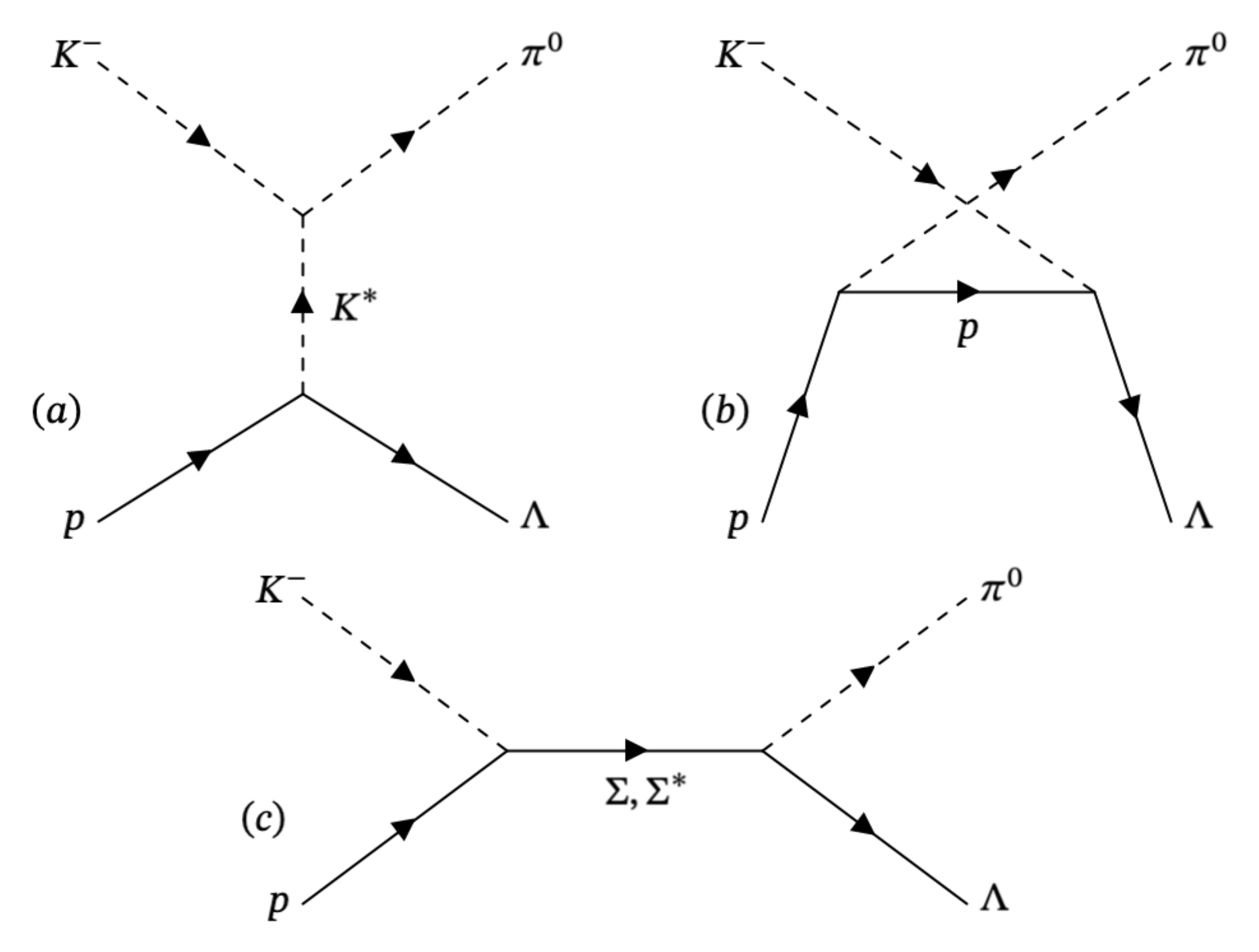}
\caption{Feynman diagrams for $K^-p\rightarrow\pi^0\Lambda$:
(a) t-channel $K^*$ exchange; (b) u-channel proton exchange;
(c) s-channel $\Sigma$ and its resonances exchanges.\label{fig:feynmandiagram}}
\end{figure}

\textit{Theoretical Background:}\label{sec:theoretical}  We employ the effective Lagrangian method which is generally used in PWA to compute the differential 
cross section and $\Lambda$ polarization of $K^-p\to \pi^0\Lambda$. We follow the theoretical framework of our previous work~\cite{Gao:2010ve,Gao:2012zh} 
and the details can be found in the Supplemental Materials~\cite{supplemental}.
The Feynman diagrams of $K^-p\rightarrow\pi^0\Lambda$ is shown in Fig.~\ref{fig:feynmandiagram}.
The building blocks of the amplitude includes the effective vertices, form factors and propagators of the exchanged particles, 
which contains cut-off parameter $\Lambda$ 
in the form-factor, coupling constants $f$, mass $M$ and width $\Gamma$ (for unstable resonances) of the exchanged particle. 
Formally, the amplitude can be expressed as a function of these parameters
\begin{equation}
   \mathcal{M}_{\lambda\lambda^\prime}=\sum_i \mathcal{M}^i_{\lambda\lambda^\prime}(\Lambda_i,~f_i, ~m_i,~\Gamma_i),
\end{equation}
where $i$ is the label for the exchanged particles, $\lambda$ and $\lambda^\prime$ stand for the spin index of proton and $\Lambda$ hyperon.
The differential cross section and $\Lambda$ polarization~\cite{Penner:2002ma} can be separately expressed as
\begin{equation}
  \frac{d\sigma}{d\Omega}=\frac{1}{64 \pi^2 s}\frac{\left|\bf{q}\right|}{\left|\bf{k}\right|}\overline{\left|\mathcal M\right|}^2,
\end{equation}
\begin{equation}
  P_\Lambda = 2\text{Im}\left(\mathcal{M}_{\frac{1}{2}}\mathcal{M}^\ast_{-\frac{1}{2}}\right){\Big{/}}\overline{|\mathcal{M}|}^2,
\end{equation}
where $d\Omega=2\pi d\cos{\theta}$ with $\theta$ the angle between the outgoing $\pi$ and the beam
direction in the c.m. frame; $s=(p+k)^2$, and $\bf{k}$ and $\bf{q}$
denote the 3-momenta of $K^-$ and $\pi$ in the c.m. frame,
respectively. And $\overline{\left|\mathcal M\right|}^2$ denotes the
spin averaged amplitude squared of the reaction. 

The basic procedure of partial wave analysis is to determine the resonance properties by comparing with experimental data. In practical analysis, 
the $t$-channel $K^\ast$, $u$-channel $p$, $s$-channel $\Sigma(1189)\frac{1}{2}^+$, $\Sigma(1385)\frac{3}{2}^+$, 
$\Sigma(1670)\frac{3}{2}^-$ and $\Sigma(1775)\frac{5}{2}^-$,
whose establishments are rated 4-star in PDG, are always included as background contributions. 
There are 14 tunable parameters for the basic 6 channels, which are 6 cut-off parameters for each channel, 2 coupling constants for $t$-channel $K^\ast$, 
3 SU(3) breaking factor for the couplings of $u$-channel $p$, $s$-channel $\Sigma(1189)\frac{1}{2}^+$ and $\Sigma(1385)\frac{3}{2}^+$, 
and the mass, width and coupling constant of $\Sigma(1670)\frac{3}{2}^-$. 
Other parameters are relatively precise and are fixed by PDG estimates. Extra $\Sigma$ resonances are added with $J^P=1/2^\pm,~3/2^\pm$ at different stages. 
Adding one additional resonance results in importing 4 parameters, which are the cut-off parameter, mass, width, and coupling constant of the exchanged particle. 

Note that this is a single-channel framework. Although multichannel analyses of $\overline{K}N$ 
interaction~\cite{Zhang:2013cua,Zhang:2013sva,Kamano:2014zba,Kamano:2015hxa,Fernandez-Ramirez:2015tfa,Matveev:2019igl,Sarantsev:2019xxm} 
are more sophisticated, we stick to a single-channel analysis because the main point of this article is to demonstrate the advantage of the application of NN in PWA, 
which can directly compare with our previous work. 
Moreover, we will show that the main error comes from the experimental uncertainties and we find no definite distinction comparing the results.
It is readily to accommodate the coupled channel effects in our NN method for further study using future experimental data. 

\textit{Partial-Wave Analysis using a Neural Network:} A NN can be used for both classification (CA) which 
gives discontinuous model predictions and regression (RE)
which is applicable for continuous model outputs.
In this study, CA corresponds to the identification
of the quantum number of the most significant $\Sigma$ resonance in the $K^-p$ scattering at different stages,
while RE accounts for the determination of the
parameters of those resonances.
We utilize a joint NN with convolutional and pooling layers and subsequent fully-connected feedforward (FCFF) layers.
The total number of tunable parameters is around 0.2 M ($2\times 10^5$).
The loss function is a weighted combination of cross entropy loss and the mean squared error loss, such that
both the CA and RE can be accommodated in one NN model.
The setup of the NN is empirical, and many other choices,
such as adding or removing hidden layers (neurons),
including FCFF only and the use of different loss functions
are also carefully checked, and the results are not significantly changed within error.

As in the conventional $\chi^2$ fitting approach, we carry out NN-based PWA for one-, two- and three-additional-resonance (1R, 2R and 3R) cases separately. 
In these sequential analyses, the previous determined probability of states are always inherited, thus in each case CA only needs to determine the 
probability of the newly added resonance possessing a specific quantum number $J^P=1/2^+,~1/2^-~,3/2^+$ or $3/2^-$. 
The training data are generated based on the effective theory formalism
discussed above. For each case,
we produce 20 M (5 M for each $J^P$ candidate) training data sets with 256 data points in each set.
The choice of 256 (128 differential cross-section and
128 $\Lambda$ polarization) is to be consistent with the experimental data~\cite{Prakhov:2008dc} used in the final prediction, 
and this number can be modified with no effort to accommodate other experiments.
In addition, we also generate 0.24 M valid data sets and 0.24 M testing data sets
for the tuning of hyper parameters and the performance measurement of the trained network, respectively.
To keep generality, in the data generation,
the parameters are chosen randomly with background parameters in the ranges listed in the Supplemental Materials~\cite{supplemental} 
and the ranges of the mass and width for the target resonance(s) are $1.44\sim 1.9$ GeV (1.44 GeV is about the threshold) and $0.01\sim 0.4$ GeV, 
respectively.
Numerically, it is found that the number of training data sets are well 
enough for the NN to learn the connection between the date sets and the significance of the resonances.

It is important to have a careful error estimation. The systematic error has two sources, one is the fluctuation of the models and training processes 
due to the random initial values, the other one is from NN performance on test sets. The first systematic error is controlled by training independently 
20 models with different initial values. The second systematic error is taken to be one minus accuracy and the relative uncertainty on the test sets for CA and RE, 
respectively. To estimate the statistical error,
4000 sets of mock data are generated
from a normal distribution determined by the central values and
uncertainties of the experimental data. And we take the
standard deviation of the 4000 model outputs to be the statistical error. The total error is the quadratic sum of all the three errors. 

All the models are trained using the ADAM algorithm with gradually decreasing learning rates.
We insert additional training epochs with artificially expanded data (AED), which
are generated on-the-fly by distorting the original training data assuming that the training data have the same errors as the experimental data.
This helps the model to learn to handle noisy data. With AED, the probability in CA from the central values of the experimental data is consistent 
with that when the uncertainties of experimental data are taken into account. 
Details of the training schemes are expatiated in the Supplemental Materials~\cite{supplemental}. 

\textit{Results and Discussion:} In the 1R analysis, the averaged CA accuracy is 98.5\% (96.9\% for $1/2^+$, 99.4\% for $1/2^-$, 99.3\% for $3/2^+$ and 98.5\% for $3/2^-$) 
on the test sets, such a high accuracy is quite satisfying. For convenience, we will use 1p, 1m, 3p and 3m to denote the $J^P$ quantum numbers in the following content. 
With experimental data, 20 models with different initial values give consistent 100\% probability of the first additional resonance to be 1p. 
The situation is not changed when the errors of the experimental data are taking into account. 
So the second systematic error is $1-96.9\%$, and the first systematic error and the statistical error are both zero, resulting in a final prediction on the first state to be 1p with 100.0(3.1)\% probability
\footnote{Note that the CA results we list here and thereafter are the mean values and standard errors. In principle the probabilities would act more like a log-normal distribution whose parameters
can be deduced from the mean values and standard error.}.
Given that the ranges of parameters when generating input data are wide, the accuracy are remarkable. This demonstrates the feasibility of our NN approach. 
The CA result that 1p is the most significant state is consistent with the $\chi^2$ fitting analysis~\cite{Gao:2010ve,Gao:2012zh} 
using the same theoretical framework, while we have a quantitative description of the significance and confidence. 
The mass of 1p in 1R case is determined to be $1.58(6)(1)(2)$~GeV, where the three errors are two systematic errors and statistical error as stated previously. 
This result is also consistent with
the $\chi^2$ fitting result (around 1.63~GeV) within errors and physically is a significant support of the 3-star $\Sigma(1660)1/2^+$ in PDG. 
The mass from our NN analysis in 1R case is relatively lower. However,
this result is under the assumption that there is only one additional resonance. In principle, the values will be modified when more states are involved, 
and we will not use the result in this 1R analysis as final predictions of 1p parameters.
Other parameters of 1p are listed in the Supplemental Materials~\cite{supplemental}. 

In the 2R analysis, the CA result of the 1R case is inherited, so the first state of 2R case is fixed to be 1p.
The averaged CA accuracy is 94.8\% (96.7\% for 1p1p, 98.4\% for 1p1m, 98.2\% for 1p3p and 86.0\% for 1p3m) on the test sets. 
The lower accuracy is understandable since adding one more resonance results in higher dimensional parameter space and the loss function becomes flatter.
The CA result shows that the second resonance can be 1m, 3p and 3m with probability $15(28)\%$,
$72(24)\%$, and $12(30)\%$, respectively. The total errors are used here and hereafter for convenience. 
This is consistent with the $\chi^2$ fitting-based PWA~\cite{Gao:2012zh}, where the most $\chi^2$ improvement comes from 1p3p. 
The most probable choice is 3p which has more than three-sigma statistical significance, while the probability of the other two are 
statistically consistent with zero within errors. So the second most significant resonance is determined to be 3p. 
The main error of the probabilities comes from the uncertainties of experimental data. It indicates that the current experimental uncertainties provide the main limitation in PWA. 
In the RE part, the masses of 1p are 1.62(11)~GeV for the 1p1m combination, 1.64(11)~GeV for 1p3p and 1.61(9)~GeV for 1p3m. 
The three values are consistent, which manifests the stability of the NN analysis.
Besides, the 1p mass in 2R case gets higher and closer to the mass of the PDG $\Sigma(1660)1/2^+$, indicating that though 
the contribution of the first 1p is dominant, adding the second resonance is also important.
The mass of the second resonance is determined to be 1.75(7)~GeV, 1.70(6)~GeV and 1.62(8)~GeV for 1m, 3p and 3m, respectively.
The rest RE results are shown in tables of the Supplemental material~\cite{supplemental}. The errors of the second state are not obviously 
larger than that of the first state as in the $\chi^2$ fitting approach, which again shows the strong resolution of our NN. 
The combined CA and RE results support the existence of the 1-star $\Sigma(1780)3/2^+$ in PDG.
 
\begin{figure}
    \centering
    \includegraphics[width = 0.4\textwidth]{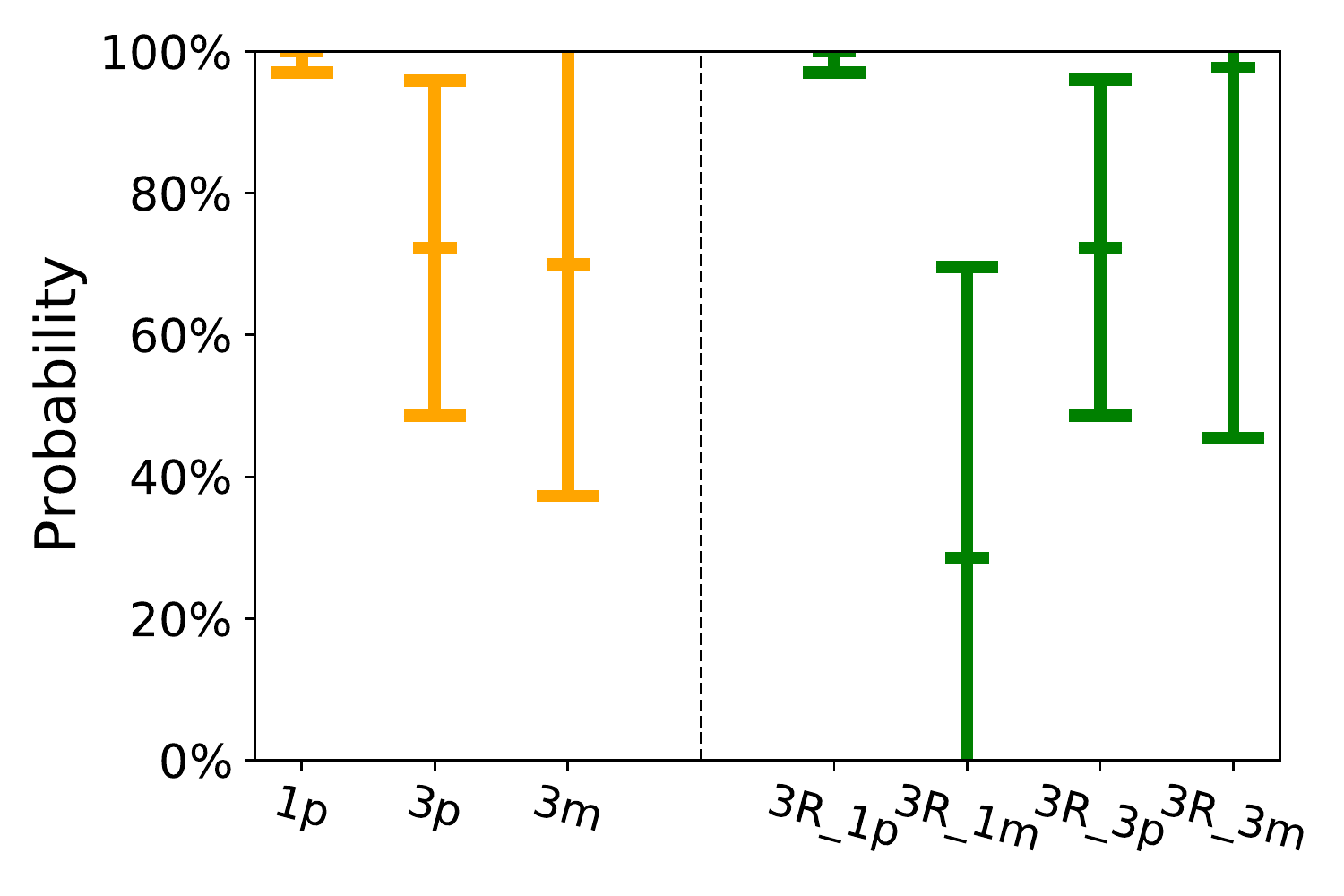}
    \caption{Probabilities of most significant resonance in the 1R, 2R and 3R case (left), and the probabilities of the four 
    candidates when assuming there are three additional resonances(right).\label{fig:CA_total}}
\end{figure}

In the 3R analysis, the CA results of the 2R and 1R case are inherited, 
so we focus on the combination 1p3pX, where ``X'' denotes the newly added resonance, that is 1p, 1m, 3p and 3m. The averaged CA accuracy in this 
case is 81.6\%. Given the first resonance to be 1p and the second one to be 3p, the third state is determined to be 3m with 97(32)\% probability, 
which is shown in the left panel of Fig.~\ref{fig:CA_total}. Accounting for the probabilities of 1p and 3p, the probability of the first three states 
being 1p3p3m is the product of the three probabilities, that is 70(33)\%. In the RE part, the masses of 1p, 3p and 3m are determined to 
be 1.62(11)~GeV, 1.72(6)~GeV and 1.62(9)~GeV, respectively. The masses of 1p and 3p are consistent with the 2R results, 
which means adding the third one does not affect the results much. On the other hand, our results support the existence of the one-star $\Sigma(1580)3/2^-$ in PDG. 
The statistical significance of the third resonance is around two sigma, thus we choose not to go beyond three additional resonances.

\begin{figure}[htbp]
 \includegraphics*[width = 0.48\textwidth]{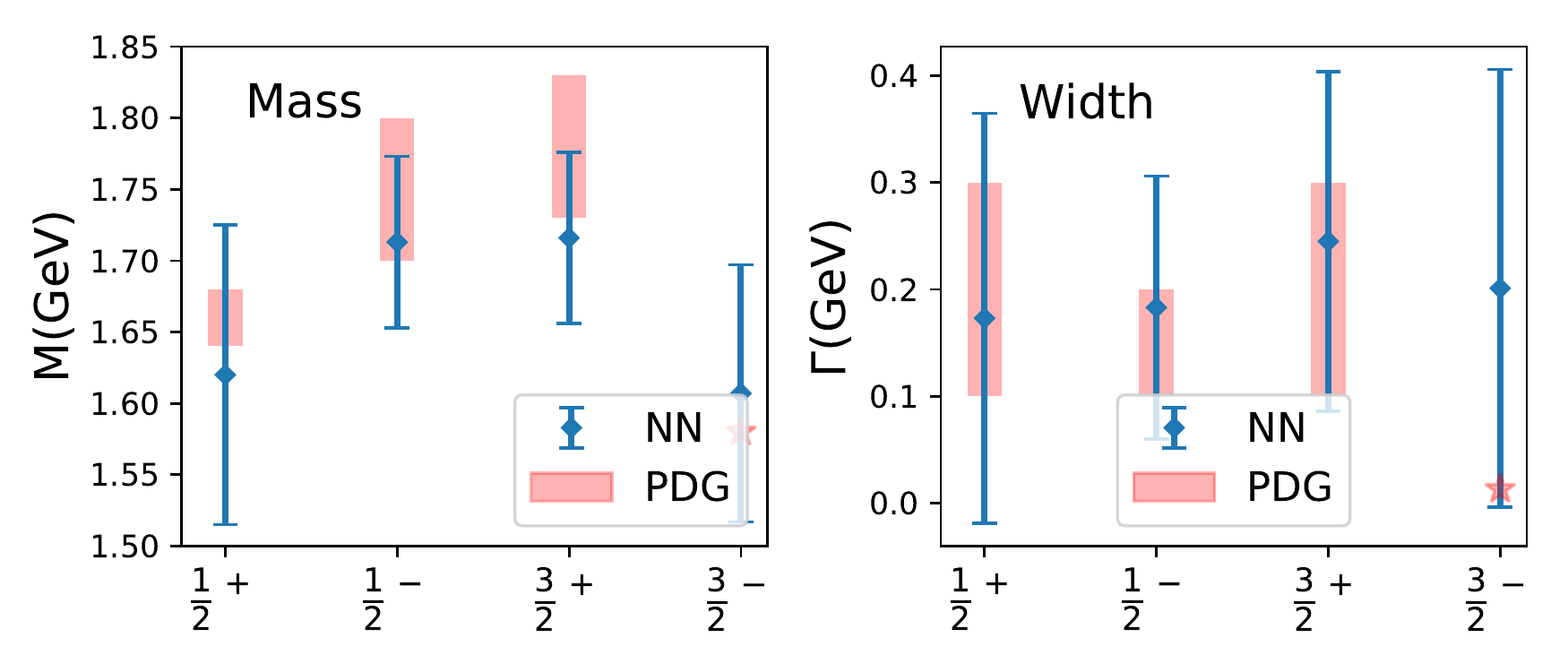}
\caption{Parameters of $\Sigma$ resonances of NN results comparing with PDG estimates.\label{fig:RE_NN_PDG}}
\end{figure}

In the above analyses, the resonance is added one by one and the probability is given separately for the first, second and third most significant resonance.
From a different point of view, on the assumption that there are three additional resonances contributing to this reaction, we can look at the probabilities of the four candidates. 
This can be done by combining the 1R, 2R and 3R probabilities, where in 3R 1p1mX and 1p3mX are also taking into account. As shown in Fig.~\ref{fig:CA_total}, 
the results are 1p with a probability of 100(3)\%, 1m with 29(41)\%, 1m with 72(24)\% and 3m with 98(52)\%. 
This four probabilities sum up to three, since we consider three additional resonances. 
And the values are consistent with the separate probabilities discussed above.
We can have a similar prediction of the properties of 1p, 1m, 3p and 3m by probability weighted average of the 
different combinations in 3R where related 2R and 1R probabilities are propagated and the results are listed in Table~\ref{tab:RE-final}.
Note that using the central values of our RE results does not guarantee a smaller $\chi^2$ than the $\chi^2$ fitting approach, 
since the RE loss function compares the parameters directly and knows nothing about $\chi^2$. The best $\chi^2$ fitting result 
should lie within uncertainties of our results.
We show the comparison of our mass and width results with the three-star $\Sigma(1660)1/2^+$, the three-star $\Sigma(1750)1/2^-$ 
and the one-star $\Sigma(1780)3/2^+$ of PDG in Fig.~\ref{fig:RE_NN_PDG}, 
which indicates that our results are consistent with PDG estimations within error. There also lists a one-star $\Sigma(1580)3/2^-$ in 
PDG with only the central value of its mass, and the mass of our 3p state is close to this resonance. 
After all our NN-based PWA analysis support the existence of the three-star $\Sigma(1660)1/2^+$, the one-star $\Sigma(1780)3/2^+$ 
and the one-star $\Sigma(1580)3/2^-$ in PDG. Our results cannot decide the existence of the three-star $\Sigma(1750)1/2^-$ since its probability is consistent with zero.

\begin{table}[htbp]
\caption{RE results for the four candidates.
\label{tab:RE-final}}
    \begin{tabular}{c|c|c|c}
    \hline\hline
~ & $m$/GeV & $\Gamma$/GeV & $\Gamma_{\overline{K} N}
  \Gamma_{\pi \Lambda} / \Gamma$\\
     \hline
$\frac{1}{2}^+$ & $1.620(105)$
  & 0.173(192) & -0.5(1.6)\\
  \hline
  $\frac{1}{2}^-$ & 1.713(60)
  & 0.183(123) & -0.60(63)\\
  \hline
  $\frac{3}{2}^+$ & 1.716(60)
  & 0.245(159) & 0.02(13)\\
   \hline
$\frac{3}{2}^-$ & 1.607(90)
  & 0.201(205) & 0.002(7)\\
  \hline\hline
\end{tabular}
\end{table}

\textit{Summary and Outlook:} 
We present a novel partial wave analysis taking advantage of neural network, which can probe the $J^P$ 
quantum number with probabilities and parameters of $\Sigma$ resonances at the same time.
With the implementation of AED technique, our NN can well handle experimental data with uncertainties and give more stable results.
The three most significant resonances are found to be $1/2^+$, $3/2^+$ and $3/2^-$ states with mass 
being 1.62(11)~GeV, 1.72(6)~GeV and 1.61(9)~GeV. 
The probabilities of the first two states have a more than 3-sigma signal, while the third one has a signal around 2-sigma.
The existence of $1/2^-$ resonance cannot get support from our analysis since its probability is consistent with zero.
The probabilities help to provide a quantitative description of the existence of $\Sigma$ resonances. The main errors come from 
the uncertainties of the experimental data, therefor reducing experimental errors will be the most efficient way in improving the situation.
Future experimental data of $\overline{K}N$ scattering are expected from J-Parc~\cite{Hicks:2012}, JLAB~\cite{KLF:2020gai} 
and the forthcoming PANDA experiment~\cite{Iazzi:2016fzb}.
Our NN-based PWA can be easily extended to accommodate other theoretical models and to include new experimental data, 
and we propose to use NN-based PWA as an alternative approach to study baryon resonances.

\begin{acknowledgments}
\section*{Acknowledgments}
We are grateful to Bing-Song Zou and Qian Wang for helpful discussions.
This work is partially supported by Guangdong Major Project of Basic and Applied Basic Research under Grant No.\ 2020B0301030008, 
Science and Technology Program of Guangzhou under Grant No.\ 2019050001.
JS is supported by
the Natural Science Foundation of China under Grant No. 12105108.
JL is supported by the Natural Science Foundation of China under Grant No.\ 12175073 and No.\ 12222503. LC is supported by the 
Natural Science Foundation of china under Grant No.\ 12175036 and the Foundation of the Hunan Provincial Education Department under Grants No.\ 20A310.
The numerical work is  done on the supercomputing system in the Southern Nuclear Science Computing Center (SNSC).
\end{acknowledgments}

\bibliographystyle{unsrt}
\bibliography{library}

\begin{widetext}





\begin{center}
{\LARGE\bf{Supplemental Materials}}\par
\end{center}

\section{Details of the Theoretical Formalism}
The details of the theoretical formalism 
constructing the amplitudes of $K^-p\to \pi^0 \Lambda$, which is used to generate the training data
for the network, is explained below. 

Relevant effective Lagrangians for hadron couplings in $K^-p\to \pi^0 \Lambda$ 
are listed in Eq.~(\ref{eq:Lagrangians_first})-(\ref{eq:Lagrangians_last}). 
\begin{eqnarray}
  \mathcal{L}_{K^*K\pi}&=&i g_{K^*K\pi}K^*_\mu(\pi\cdot\tau\partial^\mu K-\partial^\mu \pi\cdot\tau K),\label{eq:Lagrangians_first}\\
 {\mathcal L}_{K^*N\Lambda}&=&-g_{K^*N\Lambda}\overline{\Lambda}(\gamma_\mu K^{*\mu} 
  -\frac{\kappa_{K^*N\Lambda}}{2M_N}\sigma_{\mu\nu}\partial^\nu K^{*\mu})N,\\
  {\mathcal L}_{\pi NN}&=&\frac{g_{\pi NN}}{2M_N}\overline{N}\gamma^\mu\gamma_5\partial_\mu\pi\cdot\tau N,\\
  {\mathcal L}_{KN\Lambda}&=&\frac{g_{KN\Lambda}}{M_N+M_\Lambda}\overline{N}\gamma^\mu \gamma_5 \Lambda\partial_\mu K +H.c.,\\
  {\mathcal L}_{KN\Sigma(\frac{1}{2}^+)}&=&\frac{g_{KN\Sigma}}{M_N+M_\Sigma}\partial_\mu\overline{K}\overline{\Sigma}\cdot\tau\gamma^\mu\gamma_5 N+H.c.,\\
  {\mathcal L}_{\Sigma(\frac{1}{2}^+)\Lambda\pi}&=&\frac{g_{\Sigma\Lambda\pi}}{M_\Lambda+M_\Sigma} \overline{\Lambda}\gamma^\mu\gamma_5\partial_\mu\pi\cdot\Sigma + H.c.,\\
  {\mathcal L}_{KN\Sigma(\frac{1}{2}^-)}&=&-i g_{KN\Sigma}\overline{K}\overline{\Sigma}\cdot\tau N+H.c.,\\
  {\mathcal L}_{\Lambda\pi\Sigma(\frac{1}{2}^-)}&=&-i g_{\Lambda\pi\Sigma}\overline{\Sigma} \Lambda\pi + H.c.,\\
  {\mathcal L}_{KN\Sigma(\frac{3}{2}^+)}&=&\frac{f_{KN\Sigma}}{m_K}\partial_\mu\overline{K} \overline{\Sigma}^\mu\cdot\tau N + H.c.,\\
  {\mathcal L}_{\Sigma(\frac{3}{2}^+)\Lambda\pi}&=&\frac{f_{\Sigma\Lambda\pi}}{m_\pi} \partial_\mu\overline{\pi}\cdot\overline{\Sigma}^\mu\Lambda + H.c.,\\
  {\mathcal L}_{KN\Sigma(\frac{3}{2}^-)}&=&\frac{f_{KN\Sigma}}{m_K}\partial_\mu\overline{K} \overline{\Sigma}^\mu\cdot\tau\gamma_5 N + H.c.,\\
  {\mathcal L}_{\Sigma(\frac{3}{2}^-)\Lambda\pi}&=&\frac{f_{\Sigma\Lambda\pi}}{m_\pi} \partial_\mu\pi\overline{\Sigma}^\mu\gamma_5\Lambda + H.c.,\\
  {\mathcal L}_{KN\Sigma(\frac{5}{2}^-)}&=&g_{KN\Sigma}\partial_\mu\partial_\nu\overline{K} \overline{\Sigma}^{\mu\nu}\cdot\tau N+H.c.\\
  {\mathcal L}_{\Sigma(\frac{5}{2}^-)\Lambda\pi}&=&g_{\Lambda\pi\Sigma}\partial_\mu\partial_\nu \pi \cdot\overline{\Sigma}^{\mu\nu}\Lambda + H.c..\label{eq:Lagrangians_last}
\end{eqnarray}
At each vertex, the following form factor is added to describe the off-shell properties of the amplitudes as
\begin{equation}
    F(q^2, M) = \frac{\Lambda^4}{\Lambda^4 + (q^2 - M^2)^2}
\end{equation}
with $q$ and $M$ representing the cut-off parameter, 4-momentum and mass of the exchanged particle, respectively, and $\Lambda$ ranges from 0.8 to 1.5 GeV.
For the propagators with 4-momentum $q$, we use
\begin{equation}
  \frac{-g_{\mu\nu}+p^\mu p^\nu/m^2_{K^*}}{p^2-m^2_{K^*}}
\end{equation}
for $K^*$ meson exchange and the following expressions for intermediate baryons with angular momentum $J$
\begin{eqnarray}
    {J=\frac{1}{2}}&:&\frac{\not\! q +M}{q^2-M^2},\\
    {J=\frac{3}{2}}&:&\frac{\not\! q +M}{q^2-M^2}(-g^{\mu\nu}+\frac{\gamma^\mu \gamma^\nu}{3}+\frac{\gamma^\mu q^\nu-\gamma^\nu q^\mu}{3 M}+\frac{2q^\mu q^\nu}{3 M^2}),\\
    {J=\frac{5}{2}}&:&\frac{\not\! q+M}{q^2-M^2}S_{\alpha\beta\mu\nu}(q,M),
\end{eqnarray}
where
\begin{eqnarray}
  S_{\alpha\beta\mu\nu}(q,m) &=& \frac{1}{2}(\tilde{g}_{\alpha\mu}\tilde{g}_{\beta\nu}+ \tilde{g}_{\alpha\nu}\tilde{g}_{\beta\mu})-\frac{1}{5}\tilde{g}_{\alpha\beta} \tilde{g}_{\mu\nu} \nonumber\\
  & & -\frac{1}{10}(\tilde{\gamma}_\alpha\tilde{\gamma}_\mu \tilde{g}_{\beta\nu}+\tilde{\gamma}_\alpha\tilde{\gamma}_\nu \tilde{g}_{\beta\mu}\nonumber\\
  & &+\tilde{\gamma}_\beta\tilde{\gamma}_\mu  \tilde{g}_{\alpha\nu}+\tilde{\gamma}_\beta\tilde{\gamma}_\nu  \tilde{g}_{\alpha\mu}),
 \end{eqnarray}
and
 \begin{equation}
    \tilde{g}_{\mu\nu}=g_{\mu\nu}-\frac{q_\mu q_\nu}{M^2},~~~ \tilde{\gamma}_\mu
    =\gamma_\mu-\frac{q_\mu}{M^2}\not\! q.
 \end{equation}
For unstable resonances, we replace the denominator $q^2-M^2$
in the propagators by the Breit-Wigner form $q^2-M^2+iM\Gamma$, and replace $M$ in the rest of the propagators by $\sqrt{q^2}$. $M$ and $\Gamma$ in the propagators represent the
mass and total width of the exchanged resonance. With the above building blocks, 
the single-channel amplitude and hence the differential cross section and $\Lambda$ polarization expressed in the main text can be constructed.

As addressed in the main text, the t-channel $K^\ast$, u-channel proton, 
s-channel $\Sigma(1189)1/2^+$, $\Sigma(1385)3/2^+$, $\Sigma(1670)3/2^-$ and $\Sigma(1775)5/2^-$ 
are always included as background contributions with 14 tunable parameters. 
The ranges of background parameters are shown in Table~\ref{tab:param_bg}, 
where the phenomenological model predictions~\cite{Stoks:1999bz,Oh:2006hm} are used for the couplings for $K^\ast$, 
the results from SU(3) symmetry relations with adjustable factor between $1/\sqrt{2}$ and $\sqrt{2}$ are adopted 
for the couplings of u-channel proton, s-channel $\Sigma(1189)$ and $\Sigma(1385)$~\cite{Gao:2010ve}, 
while the PDG~\cite{ParticleDataGroup:2012pjm} estimations are taken for $\Sigma(1670)$.

\begin{table}[h]
\centering{}\caption{Parameters of the background, where $\Sigma^*$ denotes $\Sigma(1385)3/2^+$.\label{tab:param_bg}} 
\begin{tabular}{c|c|c|c}
  \hline
  $g_{K^{\ast} N \Lambda}$ & $g_{K^{\ast} N \Lambda} \kappa_{K^{\ast} N
  \Lambda}$ & $g_{\pi N N} g_{K N \Lambda}$ & $g_{K N \Sigma} g_{\Sigma
  \Lambda \pi}$\\
  \hline
  $\left[-6.11, -4.26\right]$ & $\left[-16.3, -10.4 \right]$ & $\left[1/\sqrt{2}, \sqrt{2}\right]\times(-176)$ 
  & $\left[1/\sqrt{2}, \sqrt{2}\right]\times 34.8$\\
  \hline\hline
  $f_{K N \Sigma^{\ast}} f_{\Sigma^{\ast} \Lambda \pi}$ & $M_{\Sigma (1670)}$ (GeV)
  & $\Gamma_{\Sigma (1670)}$ (GeV) & $\Gamma_{\overline{K} N} \Gamma_{\pi \Lambda} /
  \Gamma_{\Sigma (1670)}$\\
  \hline
  $\left[1/\sqrt{2}, \sqrt{2}\right]\times(-3.1)$ & $[1.665, 1.685]$ & [0.04, 0.1] & [0.018, 0.17]\\
  \hline
\end{tabular}
\end{table}

\section{Details of the Neural Networks}

\begin{figure}[htbp]
    \centering
    \includegraphics[width = 0.7\textwidth]{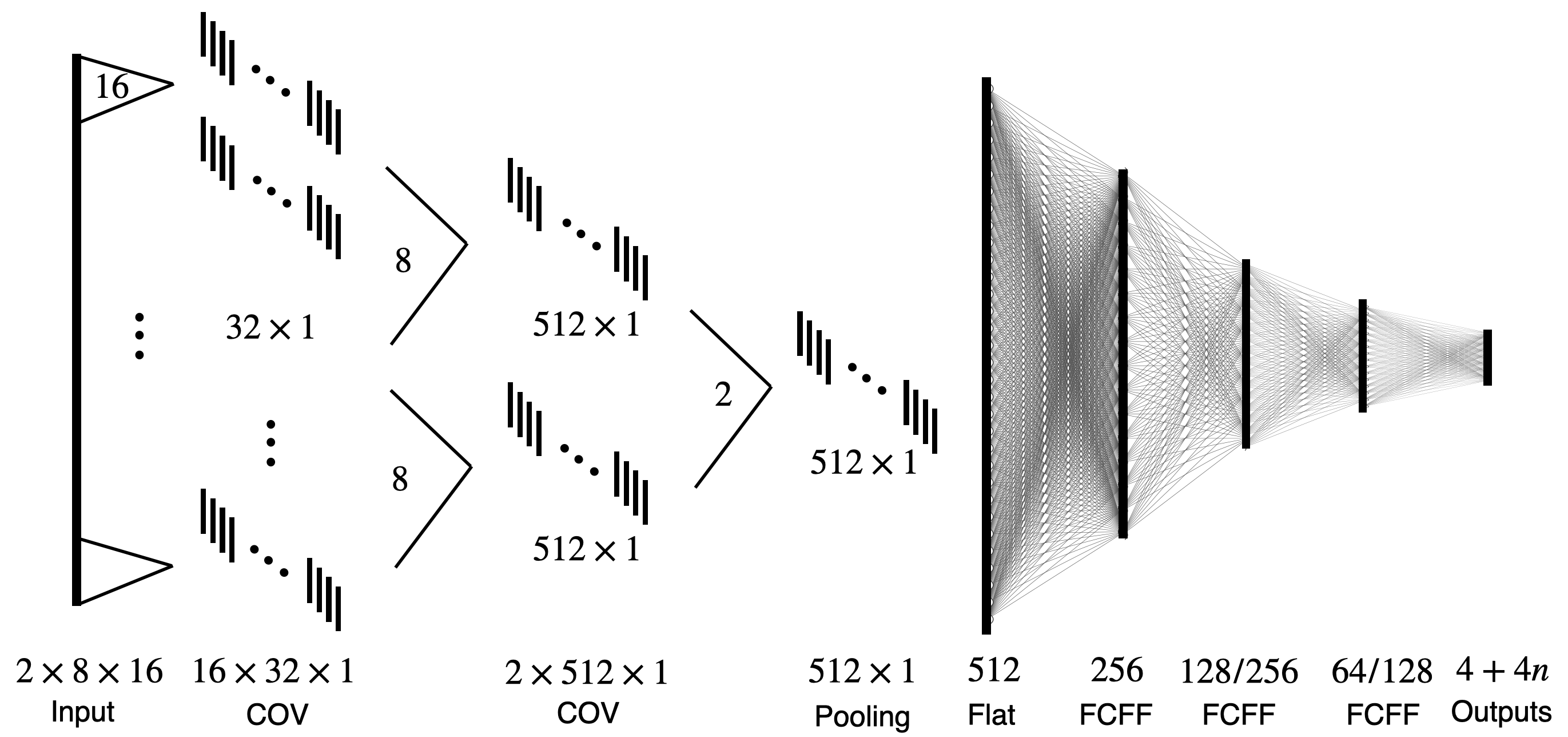}
    \caption{The NN structure used in our PWA. The number under each layer indicates the number of features or neurons.\label{fig:NN_structure}}
\end{figure}

As shown by the schematic diagram of Fig.~\ref{fig:NN_structure}, 
the neural networks (NN) are constructed empirically with 8 layers,
which is inspired by the structure of the experimental data.
Specifically, the experimental data consist of 8 energies with 16 scattering angles at each energy
for both the differential cross section and $\Lambda$ polarization measurements.
After the input layer, the first hidden layer is a convolutional (COV) layer, 
in which the data are divided into 16 groups and 32 convolutional filters are used to feature the 16 data points of each group.
We use 32 feature is to keep as much the information of the original data as possible.
Subsequently the second convolutional layer converts the data into 512 features for both the cross section data and the polarization data.
Next is the pooling layer which merges the two sets of the 512 features into one according to their importance, 
which will also be the start layer of the fully-connected feedforward (FCFF) layers.
The four FCFF layers contain 256, 128, 64 and $4+4n$ neurons (256, 256, 128 and $16$ for 3R case), 
where the first ``4'' represents four CA results in probabilities, 
$n$ indicates the number of extra resonances added and the second ``4'' denotes the four parameters (cut-off parameter, mass, width, and coupling constant) of each resonance.
In general, the convolutional and pooling layers are for extracting feature information from the experimental data, 
while the FCFF layers are for distillate the information and convert it to the desired form of outputs.

\begin{table}[h]
\centering{}\caption{Training scheme and the hyper parameters, 
where \# epochs indicates the number of epochs at each step (indicated by ``index''), 
\# AED means the proportion of AED data and the originally generated data, 
and $\alpha$ is the weight of the RE loss function.\label{tab:scheme}} 
\begin{tabular}{c|c|c|c|c}
\hline
index 	&	\# epoch 	&	\# AED & $\alpha$ &	learning rate \\
\hline\hline
1 & 10 & 1	& 0.20 & $10^{-3}$\\
\hline
2 & 10 & 1	& 0.20 & $10^{-3}$\\
\hline
3 & 10 & 1	& 0.20 & $10^{-3}$\\
\hline
4 & 10 & 1	& 0.20 & $10^{-3}$\\
\hline
5 &	10 & 1	& 0.20 & $10^{-4}$\\
\hline
6 &	10 & 1	& 0.20 & $10^{-5}$\\
\hline
\end{tabular}
\end{table}

\begin{figure}[h]
    \centering
    \includegraphics[width = 0.5\textwidth]{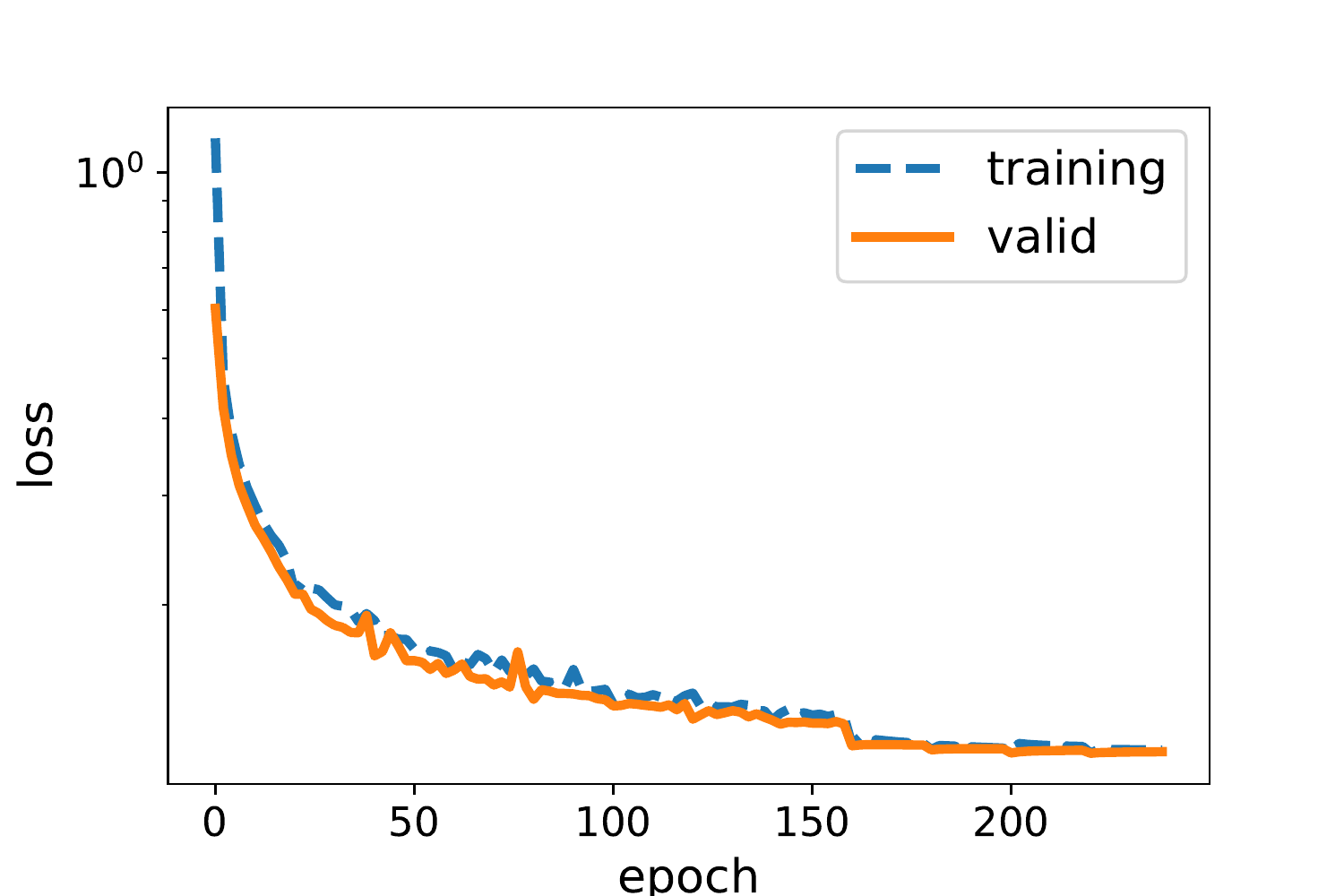}
    \caption{The training history in the 1R case.\label{fig:NN_traininghistory}}
\end{figure}

The loss function can be expressed as
\begin{equation}
    f_{\rm{loss}} = f_{\rm{CA-loss}} + \alpha f_{\rm{RE-loss}},
\end{equation}
where $f_{\rm{CA-loss}}$ and $f_{\rm{RE-loss}}$ indicates 
the cross-entropy loss function for CA and L2 loss function for RE, 
respectively, and $\alpha$ is a hyper parameter as the weight of $f_{\rm{RE-loss}}$. 
Thus both the CA and RE parts can be accommodated in one NN model.
$\alpha$ is thus chosen to make the two loss functions compatible on the valid data sets.
As pointed out in the main text, the inclusion of the artificially expanded data (AED) in the training procedure is a main improvement of our analysis. 
The choices with no AED, with the ratio AED/the original generated data being one or two in the training procedure are investigated, 
which shows that with AED data, the CA results using the central values of experimental data from softmax function and those from counting the assignment of the 4000 
sets of mock data considering experimental uncertainties are compatible, and the results with one-time and two-time AED data are nearly the same. 
Thus one-time AED data can handle the noisy data well enough and is employed in our practical analysis. 
All the models are trained using the ADAM algorithm with gradually decreasing learning rates from $1\times 10^{-3}$ to $1\times 10^{-5}$. 
The combination of those hyper parameters are called training scheme in our analysis, and the details are shown in Table~\ref{tab:scheme}. 
Please note that although the first four steps in the scheme seems the same, they are not equivalent with one step of 40 epochs since our data are read in by chunks. 
As an example, the whole training history in the 1R case is shown in Fig.~\ref{fig:NN_traininghistory}, 
from which we see that the loss on training data and valid data decrease synchronously which indicates that there is no overfitting in our NN.

There are other choices of NN set-up, such as adding or removing hidden
layers (neurons), including FCFF layers only and the use of different loss functions. 
We check the convolutional NN and FCFF NN with different number of hidden layers and training epochs, 
and find that the results are not significantly changed within errors, for example the difference of the accuracy in the 1R case is less than one percent.
So in the end we choose the convolutional NN with the above scheme to study the $\Sigma$ resonances.

\section{Additional Tables of NN PWA Results}
The CA results that are the probabilities of newly added resonance possessing 
a specific quantum number $J^P=1/2^+,~1/2^-~,3/2^+$ or $3/2^-$ in 1R, 2R and 
3R cases including 1p1mX, 1p3pX and 1p3mX are shown in Table~\ref{tab:CA}. 
The RE results of 1R and 2R cases are shown in Table~\ref{tab:RE-1R2R},
and the RE results in 3R case (1p1mX, 1p3pX and 1p3mX) are shown in Table~\ref{tab:RE_1p1m3m_1p3p3m_1p3m1m}.

\begin{table}[htbp]
\centering{}\caption{Probabilities in percentage of the newly added 
resonance with a specific quantum number $J^P=1/2^+,~1/2^-~,3/2^+$ or $3/2^-$ 
in 1R, 2R and 3R cases, where each row represents different cases, 
for example, 3R($\frac{1}{2}^+\frac{3}{2}^+$) indicates the 1p3pX case.\label{tab:CA}}
\begin{tabular}{c|c|c|c|c}
  \hline\hline
  \diagbox{cases}{candidates} & $\frac{1}{2}^+$ & $\frac{1}{2}^-$ & $\frac{3}{2}^+$ & $\frac{3}{2}^-$\\
  \hline
  1R & 100.00(3.1) & 0.00(58) & 0.00(71) & 0(1.5)\\
  \hline
  2R($\frac{1}{2}^+$) & 0.01(3.39) & 15.47(27.64) & 72.23(23.55) & 12.29(29.78)\\
  \hline
  3R($\frac{1}{2}^+\frac{1}{2}^-$) & 0.00(15.85) & 0.00(29.45) & 0.01(5.70) &
  99.99(30.90)\\
  \hline
  3R($\frac{1}{2}^+\frac{3}{2}^+$) & 1.56(17.18) & 1.26(9.68) & 0.34(28.27) &
  96.84(32.22)\\
  \hline
  3R($\frac{1}{2}^+\frac{3}{2}^-$) & 0.27(27.86) & 98.87(10.05) & 0.59(5.70) &
  0.27(23.42)\\
  \hline\hline
\end{tabular}
\end{table}

\begin{table}[htbp]
\centering{}\caption{RE results for 1R and 2R cases, 
where each row represents different resonance in each case, 
for example, $\frac{1}{2}^-$(2R, $\frac{1}{2}^+ \frac{1}{2}^-$) means the result for the $1/2^-$ resonance in the 1p1m case.\label{tab:RE-1R2R}}
    \begin{tabular}{c|c|c|c}
  \hline\hline
 ~ & $m$/GeV & $\Gamma$/GeV & $\Gamma_{\overline{K} N}
  \Gamma_{\pi \Lambda} / \Gamma$\\
  \hline
  $\frac{1}{2}^+$(1R, $\frac{1}{2}^+$) & $1.583(66) $  & $0.155(115)$ & -0.41(80)
  \\
  \hline\hline
  $\frac{1}{2}^+$(2R, $\frac{1}{2}^+ \frac{1}{2}^-$) & 1.622(110) & 0.186(221) & -0.6(1.5)
  \\
  \hline
  $\frac{1}{2}^-$(2R, $\frac{1}{2}^+ \frac{1}{2}^-$) & 1.748(66) & 0.221(156) & -0.59(59)
  \\
  \hline\hline
   $\frac{1}{2}^+$(2R, $\frac{1}{2}^+ \frac{3}{2}^+$) & 1.636(111) & 0.185(220) & -0.6(1.4)
  \\
  \hline
  $\frac{3}{2}^+$(2R, $\frac{1}{2}^+ \frac{3}{2}^+$) & 1.705(64) & 0.224(153) & 0.023(22)
  \\
  \hline\hline
   $\frac{1}{2}^+$(2R, $\frac{1}{2}^+ \frac{3}{2}^-$) & 1.610(90) & 0.183(172) & -0.47(96)
  \\
  \hline
  $\frac{3}{2}^-$(2R, $\frac{1}{2}^+ \frac{3}{2}^-$) & 1.624(78) & 0.201(183) & 0.003(5)
  \\
  \hline\hline
   \end{tabular}
\end{table}

\begin{table}[htbp]
\centering{}\caption{The same as Table~\ref{tab:RE-1R2R} but for the 3R case.\label{tab:RE_1p1m3m_1p3p3m_1p3m1m}}
    \begin{tabular}{c|c|c|c}
     \hline\hline
~ & $m$/GeV & $\Gamma$/GeV & $\Gamma_{\overline{K} N}
  \Gamma_{\pi \Lambda} / \Gamma$\\
  \hline\hline
$\frac{1}{2}^+$(3R, $\frac{1}{2}^+ \frac{1}{2}^- \frac{3}{2}^-$) & 1.620(106)
  & 0.195(216) & -0.5(1.8) \\
  \hline
  $\frac{1}{2}^-$(3R, $\frac{1}{2}^+ \frac{1}{2}^- \frac{3}{2}^-$) & 1.685(62)
  & 0.180(129) & -0.39(42)\\
  \hline
  $\frac{3}{2}^-$(3R, $\frac{1}{2}^+ \frac{1}{2}^- \frac{3}{2}^-$) & 1.544(87)
  & 0.185(194) & 0.001(4)\\
   \hline\hline
$\frac{1}{2}^+$(3R, $\frac{1}{2}^+ \frac{3}{2}^+ \frac{3}{2}^-$) & 1.620(105)
  & 0.163(181) & -0.5(1.4)\\
  \hline
  $\frac{3}{2}^+$(3R, $\frac{1}{2}^+ \frac{3}{2}^+ \frac{3}{2}^-$) & 1.716(60)
  & 0.246(159) & 0.02(2)\\
  \hline
  $\frac{3}{2}^-$(3R, $\frac{1}{2}^+ \frac{3}{2}^+ \frac{3}{2}^-$) & 1.621(91)
  & 0.207(208) & 0.002(7)\\
  \hline\hline
$\frac{1}{2}^+$(3R, $\frac{1}{2}^+ \frac{3}{2}^- \frac{1}{2}^-$) & 1.629(110)
  & 0.191(226) & -0.6(1.7)\\
  \hline
  $\frac{3}{2}^-$(3R, $\frac{1}{2}^+ \frac{3}{2}^- \frac{1}{2}^-$) & 1.600(189)
  & 0.189(212) & 0.003(8)\\
  \hline
  $\frac{1}{2}^-$(3R, $\frac{1}{2}^+ \frac{3}{2}^- \frac{1}{2}^-$) & 1.737(59)
  & 0.185(118) & -0.78(81)\\
  \hline\hline
   \end{tabular}
\end{table}
\clearpage


\end{widetext}

\end{document}